\documentclass[a4paper,fleqn,usenatbib]{mnras}

\usepackage{savesym}
\usepackage{amsmath}
\savesymbol{iint}
\savesymbol{iiint}
\usepackage{txfonts}
\restoresymbol{TXF}{iint}
\restoresymbol{TXF}{iiint}

\usepackage[T1]{fontenc}
\usepackage{ae,aecompl}

\usepackage{graphicx}	

\title[AT Cnc age]{When does an old nova become a dwarf nova? Kinematics and age of the nova shell of the dwarf nova AT Cnc}

\author[M. M. Shara et al.]{
Michael M. Shara,$^{1}$\thanks{E-mail: mshara@amnh.edu.
This paper is respectfully dedicated to the memory of Peter Wehinger, co-discoverer of AT Cancri's ejecta, 
who selflessly devoted much of his professional life to the advancement of astronomy.}
Laurent Drissen,$^{2}$
Thomas Martin,$^{2}$
Alexandre Alarie,$^{2}$
\newauthor  and F. Richard Stephenson$^{3}$
\\
$^{1}$Department of Astrophysics, American Museum of Natural History, Central Park West at 79th Street,
New York, NY 10024-5192 USA\\
$^{2}$D\'epartement de physique, de g\'enie physique et d'optique, Universit\'e Laval, Qu\'ebec (QC), G1V 0A6, Canada\\
$^{3}$Department of Physics, Durham University, South Road, Durham DH1 3LE, UK\\
}

\date{Accepted XXX. Received YYY; in original form ZZZ}

\pubyear{2016}

\begin{document}
\label{firstpage}
\pagerange{\pageref{firstpage}--\pageref{lastpage}}
\maketitle

\begin{abstract}
The Z Cam-type dwarf nova AT Cnc displays a classical nova (CN) shell, demonstrating that mass transfer in cataclysmic binaries decreases 
substantially after a CN eruption. The hibernation scenario of cataclysmic binaries predicts such a decrease, on a timescale of a few centuries.
In order to measure the time since AT Cnc's last CN eruption, we have measured the radial velocities of a hundred clumps in its ejecta with SITELLE, CFHT's recently commissioned imaging Fourier transform spectrometer. These range from -455 to +490 km/s. Coupled with the known distance to AT Cnc of 460 pc (Shara 2012), the size of AT Cnc's shell, and a simple model of nova ejecta deceleration, we determine that the last CN eruption of this system occurred $330_{-90}^{+135}$ years ago. This is the most rapid transition from a high mass transfer rate, novalike variable to a low mass transfer rate, dwarf nova yet measured, and in accord with the hibernation scenario of cataclysmic binaries. We conclude by noting the similarity in deduced outburst date (within a century of 1686 CE) of AT Cnc with a ``guest star" reported in the constellation Cancer by Korean observers in 1645 CE.

\end{abstract}

\begin{keywords}
stars: individual: AT Cnc -- (stars:) novae, cataclysmic variables -- techniques: imaging spectroscopy -- ISM: jets and outflows
\end{keywords}



\section{Introduction}

Dwarf novae (DN) and classical novae (CN) are all close binary stars, wherein a white dwarf (WD) accretes hydrogen-rich matter from its red dwarf (RD), Roche lobe-filling companion, or from the wind of a nearby giant. In DN, a thermal instability episodically dumps much of the accretion disk onto the WD \citep{1974PASJ...26..429O}. The liberation of gravitational potential energy then brightens these systems by up to 100-fold as DN eruptions occur, typically every few weeks or months. This accretion process in DN must inevitably build an electron degenerate, hydrogen-rich envelope on the white dwarf \citep[]{1986ApJ...311..163S}. Theory and detailed simulations predict that once the accreted mass reaches of the order of 10$^{-5}$ M$_\odot$, a thermonuclear runaway (TNR) will occur in the degenerate layer of accreted hydrogen \citep{1972ApJ...176..169S, 1978A&A....62..339P}. The TNR causes the rapid rise to $\sim 10^5$ L$_\odot$ or more, and the high-speed ejection of the accreted envelope in a classical nova explosion. 

The longterm evolution of cataclysmic binaries (CB) is driven by the mass accretion rate between nova eruptions. If that rate decreases by one or more orders of magnitude in the centuries following a CN eruption, as predicted by the hibernation scenario of cataclysmic binaries \citep{1986ApJ...311..163S,1986ApJ...311..172P,1987Ap&SS.131..419K}, then the predicted relative numbers of long and short period CB, the relative numbers of DN and novalike binaries, and the lifetimes of CBs all change dramatically. Population synthesis codes that model CB and their effects on the chemical evolution of galaxies \citep{2003A&A...405...23M,2016MNRAS.458.2916C} can only provide realistic predictions if accurate mass transfer rate histories are an inherent part of those codes.  

The accretion rates onto the WDs in DN are typically 10-100 times smaller than those observed in the WD-RD, mass-transferring binaries known as novalike variables. 
\citet{2009AJ....138.1846C} demonstrated that almost all pre- and post-nova binaries, observed in the century before or after eruption, are high mass-transfer rate novalike variables.  Thus the discovery of a CN shell, almost one degree in diameter, surrounding the prototypical dwarf nova Z Cam \citep[]{2007Natur.446..159S} was unexpected, and has important implications for our understanding of the longterm evolution of CBs. The derived shell mass of Z Cam matches that of classical novae, and is inconsistent with the mass expected from a dwarf nova wind or a planetary nebula. The Z Cam shell observationally linked, for the first time, a prototypical DN with an ancient CN eruption and the CN process. This was the first-ever confirmation of a key prediction of CB TNR theory: the accreting white dwarfs in DN must eventually erupt as CN. It also demonstrated that, 1300-2100 years after its CN eruption \citep{2012ApJ...756..107S}, Z Cam's central binary is {\it not} a novalike variable. Instead, it exhibits DN eruptions, indicative of a lower mass-transfer rate than is seen in old novae up to one century after eruption. The hibernation scenario of CBs predicts that the transition from high to low mass transfer state in old novae is just the WD cooling time after a CN eruption - a few centuries. A nova that erupted more recently than Z Cam is required to more stringently test the predicted transition timescale. 

Thus motivated, we have been searching for other CN shells surrounding DN. One of our targets was the Z Cam-like DN AT Cancri. Short and long-term variability in the spectrum of AT Cnc has been well documented by \citet{1999PASJ...51..115N} and references therein, who detected a clear radial velocity variation with a period of 0.2011d, a semi-amplitude of 80 km$^{-1}$ and a systemic velocity of 11 km$^{-1}$. Superhumps were detected by \citet{2004A&A...419.1035K}, who suggested that AT Cnc has a large mass ratio, and may host a magnetic white dwarf.

Optical [NII] narrowband imaging of AT Cnc \citep[]{2012ApJ...758..121S} revealed highly fragmented rings, about 3 arcmin in diameter, surrounding the star. The spectrum of one of the brightest blobs in the ejecta is dominated by lines of [NII], [OII] and [OIII]; oxygen and nitrogen are the products of a nova TNR. The geometry of the rings suggests that we are looking at an hourglass-shaped ejection reminiscent of that of other old novae such as HR Del \citep{2003MNRAS.344.1219H}. 

We present in this paper a kinematical analysis of a hundred emission-line blobs around AT Cnc, based on a hyperspectral datacube obtained with the Canada-France-Hawaii Telescope's newly commissioned imaging Fourier Transform Spectrometer (iFTS), SITELLE. We briefly describe SITELLE in section 2, and the observations of AT Cnc and their reductions in section 3. The results are presented, and the time since the last CN eruption of AT Cnc is deduced, in section 4. We summarize our results in section 5. The spectrum of AT Cnc, and a discussion of its possible association with the Korean "guest star" of 1645 CE are discussed in appendices. 


\section{Description of the instrument}

As the instrument used to collect the data presented in this paper is new and unusual, and since this is one of the first science papers using SITELLE data, a brief introduction to the instrument is given here. SITELLE\footnote{http://cfht.hawaii.edu/Instruments/Sitelle/} is essentially an imager, at the core of which a Michelson interferometer is inserted in order to modulate the incoming light and extract the spatially-resolved spectral content of its target. The interferometer consists of a beamsplitter which separates the incoming beam into two components, one of which is sent to a fixed mirror and the other one to a moving mirror. The two beams are reflected back by these mirrors and interfere. The interferometric images are then recorded on two e2v 2048 $\times$ 2048 pixel CCDs. This is enabled by an unusual mirror configuration permitting one of the two output beams to be reflected at an angle with respect to the incoming beam (in a standard Michelson interferometer, this beam is reflected back to the source).  Moving one of the mirrors introduces an optical path difference (OPD) between the two arms of the interferometer, modulating the light according to the spectral content of the source at each location in the field of view. A series of short (10 - 100 s) images are acquired by both CCDs at different regularly spaced mirror positions, resulting in two complementary interferometric cubes which are then combined. Each pixel (0.32$''$) in the image plane records an interferogram which, after proper data processing, is transformed into a spectrum. The net result of a SITELLE observation is thus a spatially-resolved spectral cube of the target, with a field of view of 11 arcminutes and a seeing-limited spatial resolution sampled at 0.32$''$.

The cube's spectral resolution depends on the maximum optical path difference between the mirrors, and therefore the number of mirror steps, which is tailored to the observer's needs. Typical spectral resolutions vary from R $\sim$ 500 to R $\sim$ 2000, although the instrument is capable of reaching R $\sim$ 10000. It has been tested and shown to work on astronomical sources up to R $\sim$ 5000. In order to obtain the desired spectral resolution in a reasonable amount of time while reducing the background photon noise, filters are inserted in the optical path. More details about SITELLE, its metrology system and its performances, are given in \citet{2012SPIE.8446E..0UG,2014AdAst2014E...9D,2014SPIE.9147E..3ZB,Drissen2016}.

\section{Observations and data reduction}
AT Cnc was observed on 9 January, 2016, as part of SITELLE's Science Verification phase using the 
SN3 filter (flat transmission above 95\% in the 651 - 685 nm range) at a spectral resolution of R=1300. To reach this resolution, 232 equally spaced mirror steps of 2881 nm each were necessary, with an integration time of 57s per step and an overhead (CCD readout and mirror displacement) of 3.8s per step, for a total observation time of 3.9 hours. The sky was moonless and photometric, with an average seeing of 0.95$''$. The data were reduced using ORBS, SITELLE's dedicated data reduction software \citep[]{2015ASPC..495..327M,Martin2016b}.

\begin{figure}
	\includegraphics[width=\columnwidth]{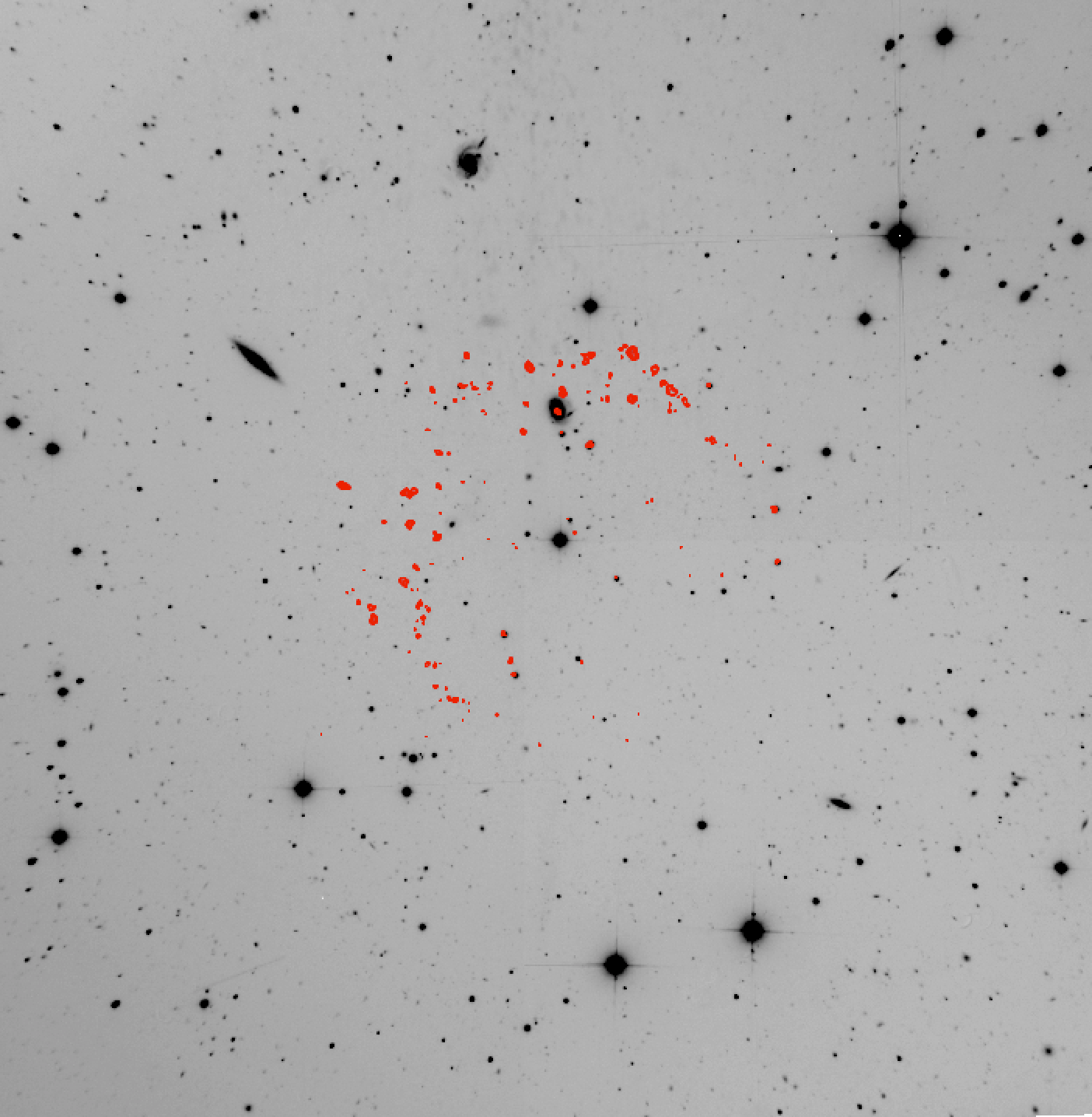}
    \caption{Deep image of the entire field of view, with contours (in red) of the most prominent emission-line blobs around AT Cnc. The field of view is $11' \times 11'$, with north at the top and east to the left. Numerous galaxies are seen in the field, some of which having redshifts that put their emission lines in the SN3 filter. They will be discussed in \citet{Drissen2016}.}
    \label{fig:Deep}
\end{figure}

Fig.~\ref{fig:Deep} shows the deep image of the field, obtained by summing all interferograms. As was already noted in \citet[]{2012ApJ...758..121S}, the ejecta are very asymmetrically distributed about AT Cnc, with the majority of the blobs located to its northwest, north, east and southeast. Almost no blobs are seen to AT Cnc's southwest. A possible explanation for this asymmetry is connected to the observation that only $\sim$ 47\% of old novae display ejected shells \citep{1985ApJ...292...90C,2000AJ....120.2007D}. Shock interaction with the surrounding interstellar medium (ISM) is essential to excite the forbidden nitrogen lines that provide the strongest emission lines seen in most old nova ejecta. Novae with no detectable ejected shells may simply lie in regions of low ISM. If the ISM surrounding AT Cnc is significantly denser to the north and east then the asymmetry in the ejecta is easily understood.

\smallskip

\subsection{Wavelength and flux calibration}
\begin{figure}
	\includegraphics[width=\columnwidth]{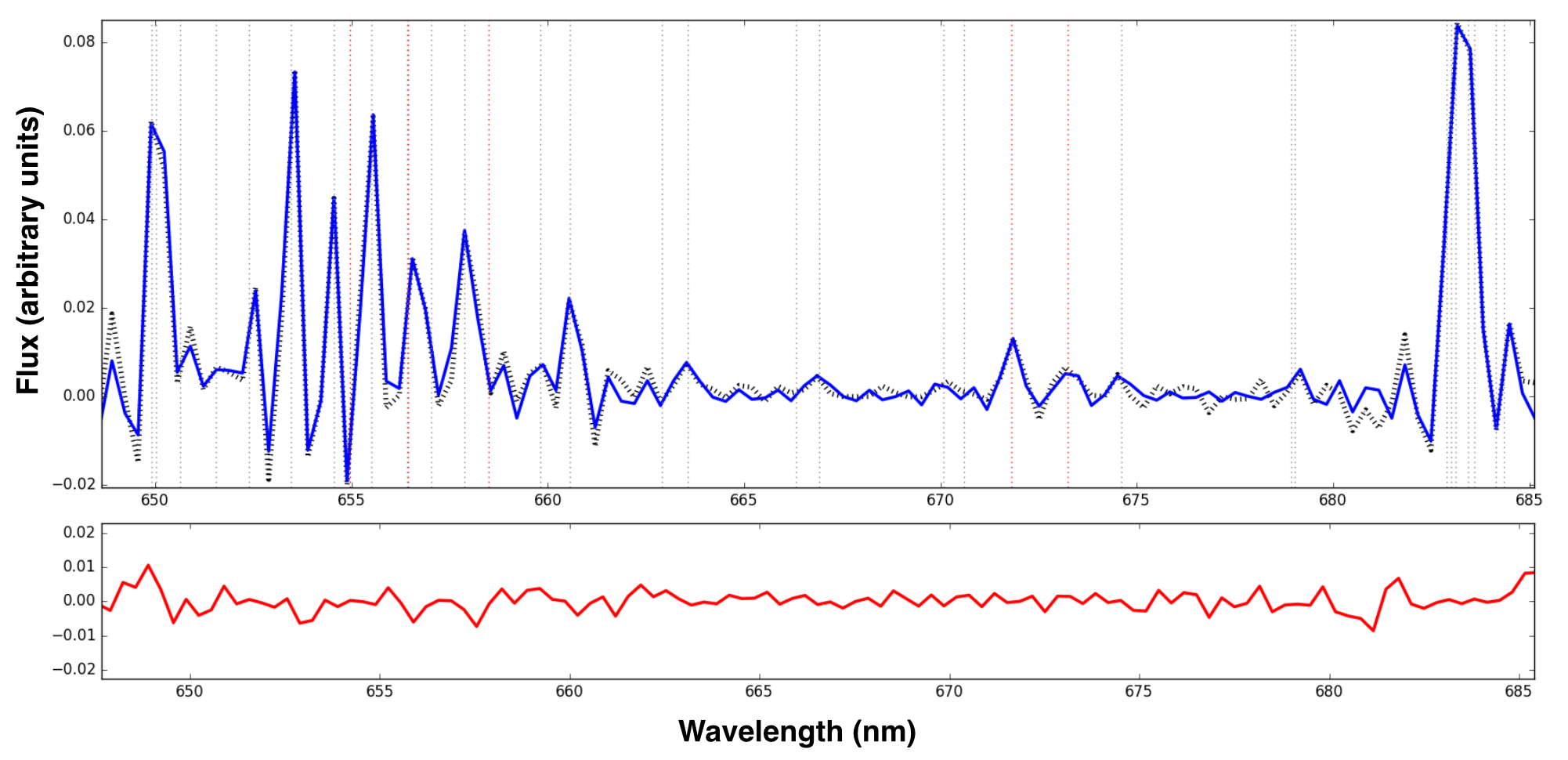}
    \caption{Average observed spectrum of the night sky (black) and fit to the OH lines (upper panel); residuals are shown in the lower panel.}
    \label{fig:night-sky}
\end{figure}

Wavelength calibration with SITELLE is secured using a high-resolution He-Ne (543.5 nm) data cube obtained at least once during the observing run. We have noticed small zero-point offsets from one cube to the next, caused in part by the fact that the targets are observed at an angle compared with the calibration laser cube. However, cubes obtained with the SN3 filter can be calibrated with an absolute precision of about 1 - 3 km/s thanks to numerous night-sky OH lines (Fig.~\ref{fig:night-sky}) filling the entire field of view of the detector. Therefore, the uncertainties in the blobs' radial velocities are completely dominated by uncertainties in the line fitting (and thus the signal-to-noise ratio of individual spectra) rather than the instrumental zero-point. 

Flux calibration is obtained with regular observations of spectrophotometric standard stars during the night, tied to a datacube of the spectrophotometrically calibrated compact planetary nebula M1-71 obtained during a previous observing run. Because of possible variations of the interferometer's modulation efficiency from one night to the next, we estimate that the absolute uncertainty on the flux is of the order of 10\%.

\subsection{Blob identification and radial velocities}

\begin{figure}
	\includegraphics[width=\columnwidth]{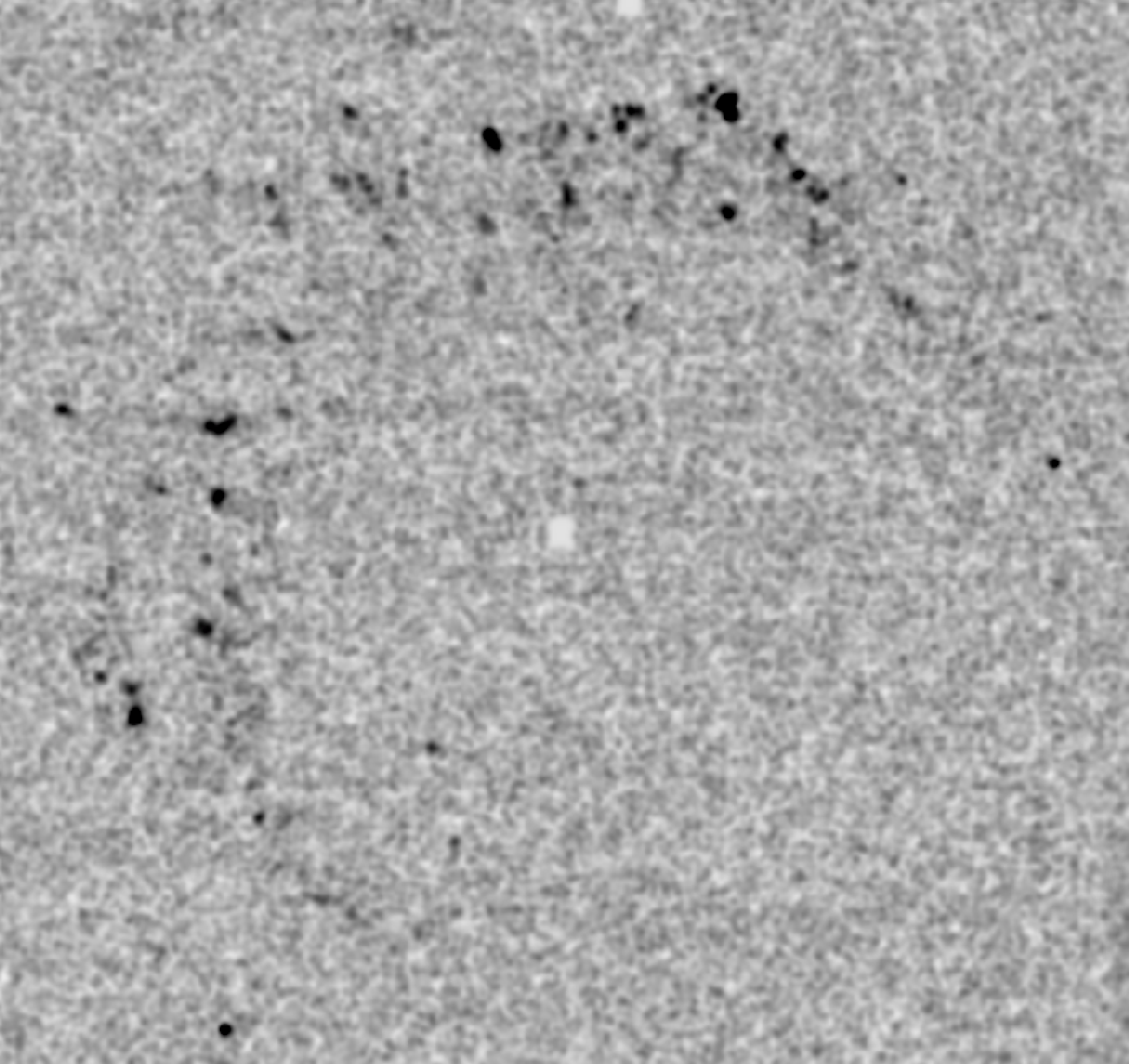}
    \caption{Frame 125 from the cube, centered at 6585 \AA. The image has been convolved with a 2.5-pixel gaussian in order to enhance the
    blobs' appearance over the background.}
    \label{fig:Frame125}
\end{figure}

Because the [NII] doublet, with its associated blueshift and redshift, fall within the night-sky OH-line forest, an average spectrum of the sky, obtained from regions in the cube without bright targets, was subtracted from the original cube before identifying the blobs.
Blobs were then visually detected and identified in individual frames (see Fig.~\ref{fig:Frame125}) of the cube. Whereas the majority of them were well defined, some were more diffuse and thus more difficult to characterize; these diffuse and faint structures are not included in the present analysis. 

\begin{figure}
	\includegraphics[width=\columnwidth]{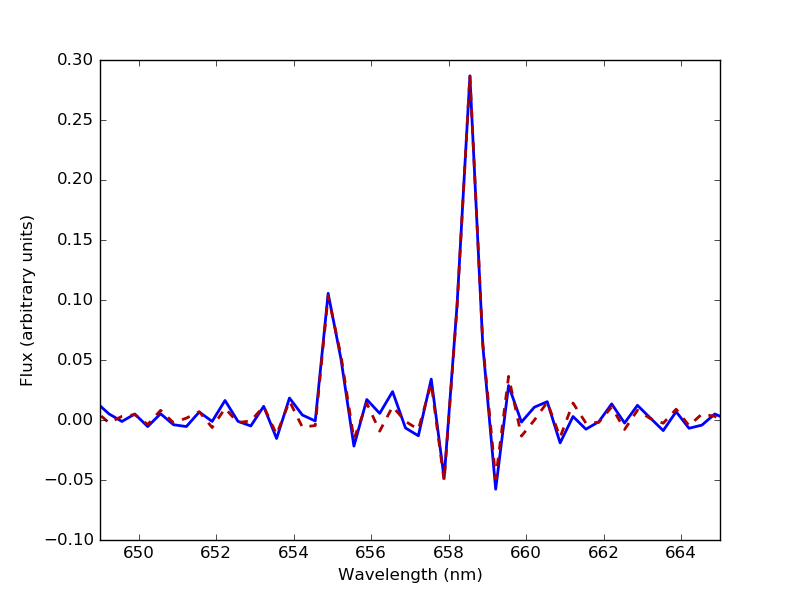}
		\includegraphics[width=\columnwidth]{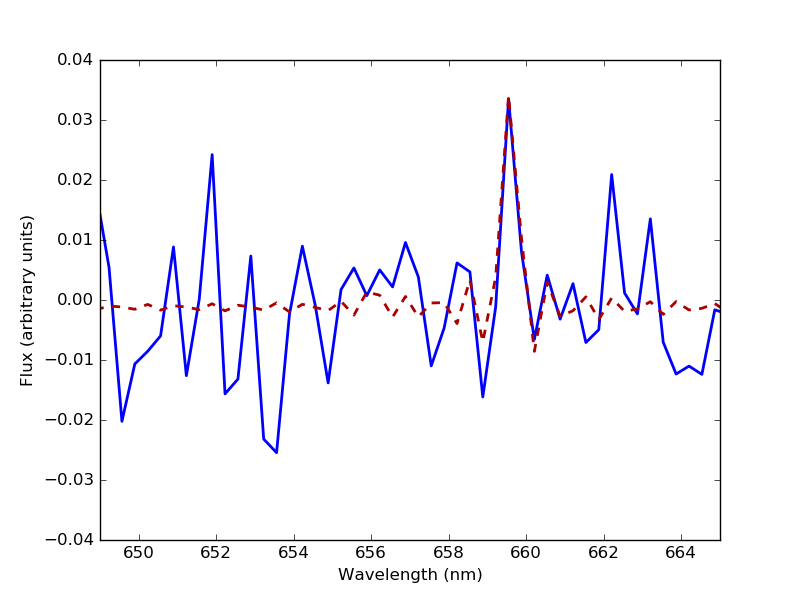}
    \caption{Spectra of blobs illustrating the quality of the fits used to determine their radial velocities in two extreme cases. Data are in solid blue, while the fits are represented by red dashed lines. The upper panel shows the spectrum of blob number 50, having the lowest uncertainty on the radial velocity (2 km/s); oscillations across the spectrum are not noise, but rather characteristic of the sinc function corresponding to an FTS Instrument line function. The lower panel shows the spectrum of blob number 114, with the [NII] 6584 line barely above the noise; it has the largest uncertainty (28 km/s).}
    \label{fig:Blobsfit}
\end{figure}

Spectra were extracted with SAOImage DS9, using the most appropriate aperture given the shape and orientation of each blob. A fitting routine using the correct instrument line shape (essentially a convolution of a sinc function with a gaussian) described in \citet{Martin2016} was then applied to each spectrum, providing radial velocities and reliable uncertainties (Tab.~\ref{tab:Blobs_table}, where blobs are listed in increasing order of radial velocity, from the most blueshifted to the most redshifted). In 51 cases, the fainter [NII] 6548 line was clearly detected and fitted. Examples of spectra and fits are shown in Fig.~\ref{fig:Blobsfit}, for the brightest and the faintest blobs in our sample. 
Uncertainties on radial velocities range from 2 km/s to 28 km/s (clearly anti-correlated with the flux in the [NII] lines, as expected), with an average of 12 km/s, which is less than a tenth of a velocity channel and about 1\% of the global velocity range ($\sim 1000$ km/s) in the shell.

\section{The Expansion and Age of the AT Cnc Ejecta}
\subsection{Expansion}

\begin{figure}
	\includegraphics[width=\columnwidth]{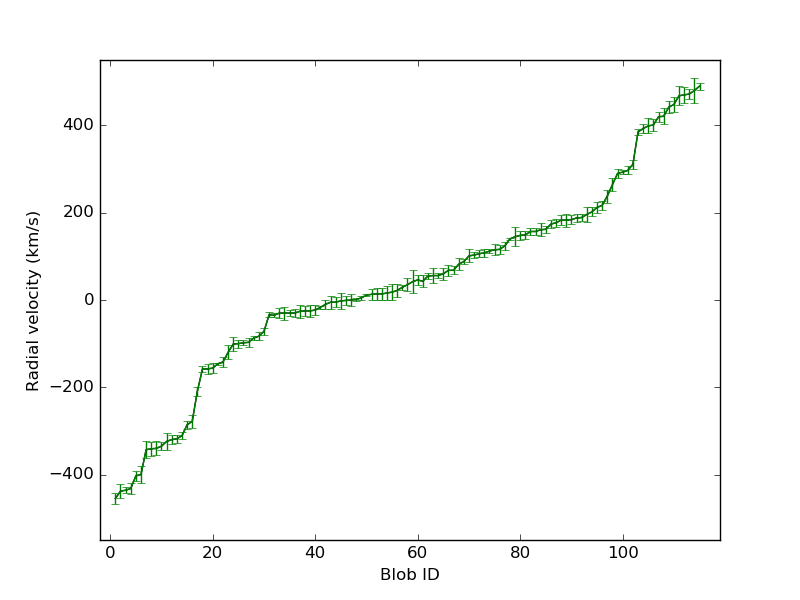}
    \caption{Distribution of radial velocities. Note the sharp discontinuities between -160 and -275 km/s, and between +310 and +400 km/s.}
    \label{fig:RVorder}
\end{figure}

In Fig.~\ref{fig:RVorder}, we plot the distribution of radial velocities in order of increasing values. Two sharp discontinuities are seen, 
between -160 and -275 km/s, and between +310 and +400 km/s. These discontinuities are likely associated with the transition from polar caps to an equatorial ring, an ejecta geometry often seen in classical novae \citep{1999MNRAS.307..677G,2003MNRAS.344.1219H, 2012A&A...545A..63C}. 475 km/s is the maximum projected ejection velocity observed with respect to the central star. Since the distance to AT CnC is reasonably well established at $460_{-133}^{+186}$ pc \citep[]{2012ApJ...758..121S}, so is the largest projected distance of an ejecta blob, on the equatorial belt:  $0.22_{-0.06}^{+0.09}$ pc. 

\begin{figure}
	\includegraphics[width=\columnwidth]{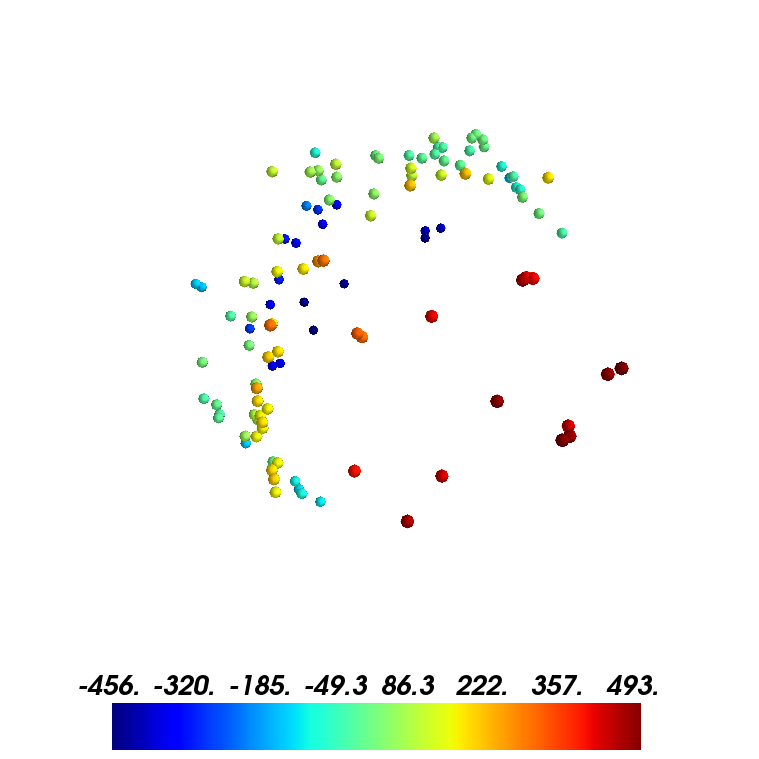}
    \caption{Exerpt from the three-dimensional animation, showing the AT Cnc nebula in the ($\alpha$, $\delta$) plane, color-coded according the
    blobs' radial velocities. Refer to Movie 1 for an animation
of these data (available in the online journal). }
    \label{fig:Figvideo1}
\end{figure}

\begin{figure}
	\includegraphics[width=\columnwidth]{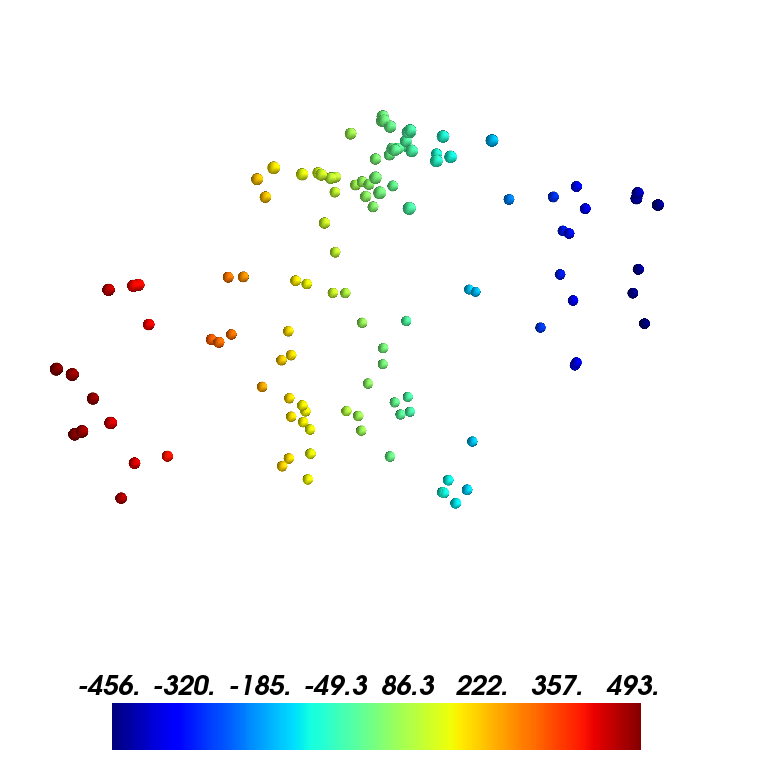}
    \caption{Exerpt from the three-dimensional animation, showing the AT Cnc nebula in the (RV, $\delta$) plane, color-coded according to the
    blobs' radial velocities. The observer is located on the right. Refer to Movie 1 for an animation
of these data (available in the online journal).}
    \label{fig:Figvideo2}
\end{figure}

If we assume a linear scaling factor between observed radial velocities of blobs and projected distance from the plane of the sky containing AT Cnc, we can model the 3D distribution of the blobs. A value of 1 km/s = 1 pixel yields the 3D distribution of the blobs depicted in Fig.~\ref{fig:Figvideo1} and Fig.~\ref{fig:Figvideo2}.
While Figure 7 supports the suggestion that the discontinuities noted in Figure 5 above are, in fact, associated with the transition from polar caps to an equatorial ring,   the scaling factor adopted above may be larger or smaller than the value adopted, expanding or shrinking the elongation of the polar blobs with respect to the equatorial ring. Figures 6 and 7 and Movie 1 suggest that the axis joining the polar blobs is inclined at $\sim 30-40$ deg to our line of sight, so that the largest radial velocities, with respect to AT Cnc, in the polar caps would then be 15-30\% larger than the measured value of 475 km/s, i.e. $\sim$ 550-600 km/s. 

\subsection{Age}

Since we can't precisely correct for projection effects, either of the radial velocity or distance from AT Cnc of the most distant blobs, the simplest and most conservative approach is to adopt the maximum observed extension ($0.22_{-0.06}^{+0.09}$ pc) and the maximum observed radial velocity (475 km/s) as we estimate the time T since AT Cnc's last nova eruption. (The age estimated below increases by only a few decades if we adopt velocities and maximal distances corrected for inclination, as these tend to cancel each other). If we ignore deceleration of the ejecta we find T = $489_{-133}^{+200}$ yrs. 

However, nova shells ${\it are}$ observed to decelerate, with ejection velocities halving on a timescale of 75 yr \citep{1946MNRAS.106..159O, 1987Ap&SS.131..461D}. During the initial, free-expansion phase after a nova eruption (with a time t ranging from a few years to a few decades), the ejected mass is much larger than the mass swept up from the interstellar medium. Thus the ejecta expand in radius R at nearly constant velocity v, so that R $\sim$ t. Once the swept-up mass is larger than the ejecta mass (the ``Sedov-Taylor" phase), deceleration occurs, so that for constant interstellar matter density one finds R $\sim t^{0.4}$ and v $\sim t^{-0.6} \sim R^{-1.5}$ \citep{1950RSPSA.201..159T}. These relationships yield an initial ejection velocity from AT Cnc of 2200km/s that occurred $330_{-90}^{+135}$ years ago. We note again that this age barely changes if we adopt inclination-corrected velocities and maximal distances (as larger velocities and distances cancel each other).
This remarkably short timescale is in excellent agreement with the prediction of the decline time of mass transfer after a nova eruption \citep{1987Ap&SS.131..419K}. The high deduced initial ejection velocity implies that AT Cnc was an intrinsically luminous, fast nova \citep{2005ApJ...623..398Y}.

As we've already noted, this young age has important implications for our understanding of the longterm evolution of CBs. While most novae which erupted in the past century are now ``novalike variables", \citep{2009AJ....138.1846C}, transferring matter at high rates from their red dwarfs to their white dwarf primaries, only a handful of older novae have been identified. 233 years after it erupted, WY Sge (nova Sge 1783) \citep{1983ApJ...264..560S,1984ApJ...282..763S} is still transferring mass at a substantial rate \citep{1996MNRAS.278..845S}. Dwarf nova behavior (and hence lower mass accretion rate) is observed in the two-thousand year old novae Z Cam \citep{2007Natur.446..159S, 2012ApJ...756..107S} and BK Lyn \citep{2013MNRAS.434.1902P}. There has not been, until now, a nova which ``bridged the gap" between two centuries and two millennia. AT Cnc is now established as that nova. It demonstrates that, 240 - 465 years after it erupted as a CN, it has become a low mass-transfer rate DN.

Finally, we note that the deduced eruption date - 1686 CE, with an error of roughly 1 century in either direction, is in remarkable agreement with a ``guest star" reported by Korean observers in 1645 CE in the asterism Yugui (part of the constellation Cancer) \citep{1966Sci...154...597}. We discuss whether this 1645 CE transient might have been the nova outburst of AT Cnc, which generated the ejecta we discuss in this paper, in an appendix below.

\begin{table}
	\centering
	\caption{Properties of blobs in AT Cnc}
	\label{tab:Blobs_table}
	\begin{tabular}{lllrrc} 
		\hline
		ID & RA & Dec & RV&RV error& Flux \\
		     &J2000.0& J2000.0 & (km/s) & (km/s) & (10$^{-16}$ erg/s/cm$^2$)\\
		\hline
1	&	8:28:41.37 &	+25:20:10.6	&	-455	&	13	&	2.4	\\
2	&	8:28:35.88 &     +25:21:09.8&	-438	&	16	&	1.4	\\
3	&	8:28:41.78 & +25:20:28.8	&	-436	&	7	&	4.3	\\
4	&	8:28:39.83&  +25:20:40.4	&	-432	&	12	&	2.8	\\
5	&	8:28:35.90 & +25:21:13.4	&	-403	&	12	&	2,6	\\
6	&	8:28:35.15 & +25:21:14.9	&	-400	&	19	&	1.6	\\
7	&	8:28:40.75 & +25:21:16.6	&	-343	&	19	&	0.9	\\
8	&	8:28:42.82 & +25:19:48.5	&	-341	&	16	&	1.6	\\
9	&	8:28:43.18 & +25:19:46.9	&	-340	&	16	&	1.3	\\
10	&	8:28:43.24 & +25:20:25.8	&	-335	&	9	&	3.2	\\
11	&	8:28:40.07 & +25:21:28.1	&	-324	&	19	&	1.0	\\
12	&	8:28:41.98 & +25:21:04.2	&	-320	&	10	&	5.2	\\
13	&	8:28:42.48 & +25:21:06.5	&	-311	&	8	&	4.1	\\
14	&	8:28:42.77 & +25:20:41.0	&	-318	&	9	&	1.8	\\
15	&	8:28:40.91 & +25:21:23.7	&	-286	&	9	&	3.3	\\
16	&	8:28:44.09 & +25:20:10.1	&	-278	&	15	&	2.5	\\
17	&	8:28:41.34 & +25:21:23.2	&	-209	&	10	&	2.7	\\
18	&	8:28:45.93 & +25:20:33.8	&	-158	&	6	&	4.9	\\
19	&	8:28:32.36 & +25:21:36.1	&	-158	&	11	&	1.5	\\
20	&	8:28:44.07 & +25:19:00.7	&	-156	&	12	&	1.4	\\
21	&	8:28:46.15 & +25:20:35.4	&	-146	&	2	&	15.6	\\
22	&	8:28:41.75 & +25:18:33.4	&	-142	&	12	&	1.7	\\
23	&	8:28:40.80 & +25:18:26.4	&	-120	&	16	&	1.0	\\
24	&	8:28:41.87 & +25:18:38.9	&	-101	&	17	&	0.6	\\
25	&	8:28:41.56 & +25:18:31.7	&	-100	&	9	&	1.4	\\
26	&	8:28:32.22 & +25:21:27.6	&	-98	&	6	&	2.5	\\
27	&	8:28:41.59 & +25:18:32.1	&	-97	&	11	&	4.0	\\
28	&	8:28:40.79 & +25:21:48.7	&	-87	&	9	&	4.4	\\
29	&	8:28:32.86 & +25:21:39.4	&	-83	&	5	&	4.5	\\
30	&	8:28:32.11 & +25:21:25.3	&	-72	&	8	&	3.6	\\
31	&	8:28:32.48 & +25:21:31.2	&	-34	&	6	&	3.4	\\
32	&	8:28:44.88 & +25:19:18.7	&	-35	&	5	&	2.7	\\
33	&	8:28:35.60 & +25:21:48.4	&	-30	&	12	&	2.5	\\
34	&	8:28:34.29 & +25:21:46.2	&	-30	&	15	&	1.1	\\
35	&	8:28:30.48 & +25:20:58.8	&	-30	&	7	&	2.2	\\
36	&	8:28:45.52 & +25:19:27.9	&	-30	&	9	&	1.6	\\
37	&	8:28:35.43 & +25:21:48.0	&	-26	&	15	&	1.6	\\
38	&	8:28:36.82 & +25:21:43.7	&	-25	&	12	&	2.2	\\
39	&	8:28:44.35 & +25:20:14.4	&	-25	&	14	&	1.3	\\
40	&	8:28:35.75 & +25:21:44.2	&	-23	&	11	&	3.0	\\
41	&	8:28:44.89 & +25:19:17.1	&	-17	&	4	&	5.2	\\
42	&	8:28:36.30 & +25:21:41.4	&	-10	&	11	&	2.1	\\
43	&	8:28:34.73 & +25:21:36.8	&	-5	&	15	&	1.5	\\
44	&	8:28:44.93 & +25:19:24.6	&	-5	&	12	&	3.9	\\
45	&	8:28:35.39 & +25:21:39.2	&	-2	&	18	&	1.1	\\
46	&	8:28:42.62 & +25:18:52.4	&	-1	&	10	&	1.6	\\
47	&	8:28:40.47 & +25:21:29.5	&	0	&	13	&	1.8	\\
48	&	8:28:33.73 & +25:21:46.7	&	1	&	4	&	7.6	\\
49	&	8:28:38.21 & +25:21:42.4	&	5	&	5	&	5.0	\\
50	&	8:28:33.80 & +25:21:50.0	&	11	&	2	&	19.1	\\
51	&	8:28:34.09 & +25:21:52.9	&	13	&	13	&	1.9	\\
52	&	8:28:34.26 & +25:21:51.0	&	14	&	14	&	1.2	\\
53	&	8:28:31.55 & +25:21:08.4       &	14	&	14	&	1.7	\\
54	&	8:28:43.51 & +25:19:57.4	&	16	&	17	&	0.6	\\
55	&	8:28:45.43 & +25:19:48.3	&	18	&	18	&	0.9	\\
56	&	8:28:32.23 & +25:21:17.2	&	22	&	14	&	3.4	\\
57	&	8:28:38.09 & +25:21:39.7	&	29	&	6	&	3.6	\\
58	&	8:28:40.13 & +25:21:16.8     &	35	&	15	&	1.5	\\
59	&	8:28:39.80 & +25:21:29.0	&	42	&	27	&	1.8	\\
60	&	8:28:38.31 & +25:21:19.4     &	46	&	11	&	3.3	\\
		\hline
	\end{tabular}
\end{table}

\begin{table}
	\centering
	\contcaption{Properties of blobs in AT Cnc}
	\label{tab:Blobs_table2}
	\begin{tabular}{lllrrc} 
		\hline
		ID & RA & Dec & RV&RV error& Flux \\
		     &J2000.0& J2000.0& (km/s) & (km/s) & (10$^{-16}$ erg/s/cm$^2$)\\
		\hline
61	&	8:28:43.19 & +25:19:35.9	&	43	&	14	&	1.9	\\
62	&	8:28:43.63 & +25:19:07.5	&	55	&	7	&	6.7	\\
63	&	8:28:40.54 & +25:21:32.1	&	56	&	17	&	0.7	\\
64	&	8:28:43.30 & +25:20:12.6	&	56	&	6	&	4.5	\\
65	&	8:28:43.09 & +25:19:16.4	&	60	&	14	&	0.8	\\
66	&	8:28:40.85 & +25:21:30.9	&	68	&	13	&	1.7	\\
67	&	8:28:35.86 & +25:21:48.4	&	68	&	9	&	1.5	\\
68	&	8:28:43.19 & +25:19:19.4	&	82	&	14	&	0.9	\\
69	&	8:28:43.16 & +25:20:30.4	&	88	&	5	&	9.0	\\
70	&	8:28:39.80 & +25:21:33.3&	101	&	15	&	1.1	\\
71	&	8:28:36.79 & +25:21:26.5	&	103	&	6	&	3.6	\\
72	&	8:28:42.11 & +25:20:53.5	&	106	&	8	&	5.0	\\
73	&	8:28:42.31 & +25:21:29.5	&	108	&	10	&	1.5	\\
74	&	8:28:43.45 & +25:20:30.7	&	112	&	5	&	3.8	\\
75	&	8:28:35.65 & +25:21:26.0	&	115	&	13	&	1.5	\\
76	&	8:28:38.44 & +25:21:04.8	&	116	&	10	&	2.0	\\
77	&	8:28:36.84 & +25:21:29.5	&	124	&	10	&	1.8	\\
78	&	8:28:33.83 & +25:21:22.4	&	140	&	3	&	13.7	\\
79	&	8:28:42.18 & +25:18:54.2	&	145	&	22	&	0.8	\\
80	&	8:28:42.98 & +25:19:08.3	&	148	&	10	&	1.8	\\
81	&	8:28:42.27 & +25:18:39.0	&	149	&	10	&	4.6	\\
82	&	8:28:42.08 & +25:20:34.7	&	157	&	8	&	2.0	\\
83	&	8:28:42.80 & +25:19:19.3	&	157	&	9	&	2.5	\\
84	&	8:28:42.72 & +25:19:12.8	&	161	&	15	&	1.0	\\
85	&	8:28:42.51 & +25:19:22.8	&	162	&	8	&	4.3	\\
86	&	8:28:41.05 & +25:20:35.5	&	174	&	10	&	0.8	\\
87	&	8:28:31.61 & +25:21:21.0	&	177	&	9	&	3.5	\\
88	&	8:28:42.05 & +25:19:52.6	&	183	&	12	&	4.1	\\
89	&	8:28:42.69 & +25:19:16.0	&	183	&	15	&	2.2	\\
90	&	8:28:42.33 & +25:18:51.3	&	184	&	10	&	2.1	\\
91	&	8:28:42.84 & +25:19:27.1	&	188	&	9	&	2.7	\\
92	&	8:28:42.25 & +25:20:06.9	&	189	&	9	&	2.8	\\
93	&	8:28:42.24 & +25:18:46.8	&	196	&	18	&	2.0	\\
94	&	8:28:42.39 & +25:19:49.7	&	202	&	10	&	2.4	\\
95	&	8:28:36.92 & +25:21:16.5	&	212	&	12	&	6.4	\\
96	&	8:28:34.84 & +25:21:22.0	&	217	&	11	&	3.7	\\
97	&	8:28:42.77 & +25:19:33.9	&	238	&	15	&	1.4	\\
98	&	8:28:40.38 & +25:20:37.0	&	264	&	15	&	1.3	\\
99	&	8:28:40.17 & +25:20:36.6	&	290	&	10	&	3.1	\\
100	&	8:28:42.14 & +25:20:04.9	&	293	&	4	&	10.8	\\
101	&	8:28:38.80 & +25:19:58.4	&	297	&	9	&	1.5	\\
102	&	8:28:38.96 & +25:20:00.1	&	311	&	10	&	2.2	\\
103	&	8:28:39.10 & +25:18:54.3	&	385	&	8	&	2.4	\\
104	&	8:28:33.03 & +25:20:24.6	&	393	&	10	&	2.0	\\
105	&	8:28:32.82 & +25:20:23.9	&	399	&	17	&	1.3	\\
106	&	8:28:36.35 & +25:20:06.7	&	401	&	13	&	2.0	\\
107	&	8:28:36.09 & +25:18:52.1	&	420	&	11	&	2.4	\\
108	&	8:28:31.74 & +25:19:14.6	&	421  &	18	&	0.8	\\
109	&	8:28:33.27 & +25:20:22.3	&	442	&	14	&	1.6	\\
110	&	8:28:37.31 & +25:18:32.2	&	448	&	17	&	1.8	\\
111	&	8:28:31.84 & +25:19:10.4	&	468	&	22	&	0.3	\\
112	&	8:28:30.54 & +25:19:38.4	&	470	&	18	&	2.4	\\
113	&	8:28:34.26 & +25:19:26.8	&	472	&	12	&	6.0	\\
114	&	8:28:32.13 & +25:19:08.9	&	480	&	28	&	0.6	\\
115	&	8:28:30.16 & +25:19:40.9	&	490	&	8	&	2.5	\\
\hline
	\end{tabular}
\end{table}

\section{Conclusions}
We have measured the radial velocities of 100 blobs in the ejecta of the old nova (and currently dwarf nova) AT Cnc. The fastest moving blobs
have velocities of 475 km/s. Combined with the observed angular size of the ejecta, and the distance to AT Cnc, we have determined that the CN which generated the ejecta occurred $330_{-90}^{+135}$ years ago. It is the best-determined transition time, to date, for an old nova to become a dwarf nova. It is also consistent with the cooling time of a WD after a CN eruption, and the timescale predicted by the hibernation scenario of CB for mass transfer to decline substantially are a CN eruption. 
The deduced date of AT Cnc's nova eruption (within a century of 1686 CE) is remarkably close in time and location to a ``guest star" reported by Korean observers in Cancer in 1645 CE.

\section*{Acknowledgements}
Based  on  observations  obtained  with  SITELLE,  a  joint project  of  Universit\'e  Laval,  ABB,  Universit\'e  de  Montr\'eal
and  the  Canada-France-Hawaii  Telescope  (CFHT)  which is  operated  by  the  National  Research  Council  (NRC)  of
Canada,  the  Institut  National  des  Science  de  l'Univers  of the  Centre  National  de  la  Recherche  Scientique  (CNRS)
of France, and the University of Hawaii. LD is grateful to the Natural Sciences and Engineering Research Council of
Canada, the Fonds de Recherche du Qu\'ebec, and the Canada Foundation for Innovation for funding. We thank Dave Zurek for 
help in retrieving and understanding earlier images of AT Cnc.

\addcontentsline{toc}{section}{Appendices}
\section*{Appendices}

\renewcommand{\thesubsection}{\Alph{subsection}}

\subsection{A spectrum of the dwarf nova}

The H$\alpha$ emission line of AT Cnc is sometimes observed to be in emission, sometimes in absorption, and occasionally a 
P Cygni profile is detected.
In Fig. 8 we show the spectrum of the central binary in AT Cnc, extracted from the data cube. It clearly shows H$\alpha$ in emission, about twice as strong as the nebular lines in the blobs; after correction for the instrumental line function, it corresponds to a FWHM of $\sim$ 400 kms$^{-1}$, centered on a velocity of +24 kms$^{-1}$. The modest FWHM suggests that we are seeing the accretion disk nearly face-on, consistent with our interpretation of the highest velocity ejected blobs belonging to polar caps. There is also a hint of the He~I $\lambda$6678 in emission, as well as a broad absorption at 650 nm.

\begin{figure}
  \includegraphics[width=\columnwidth]{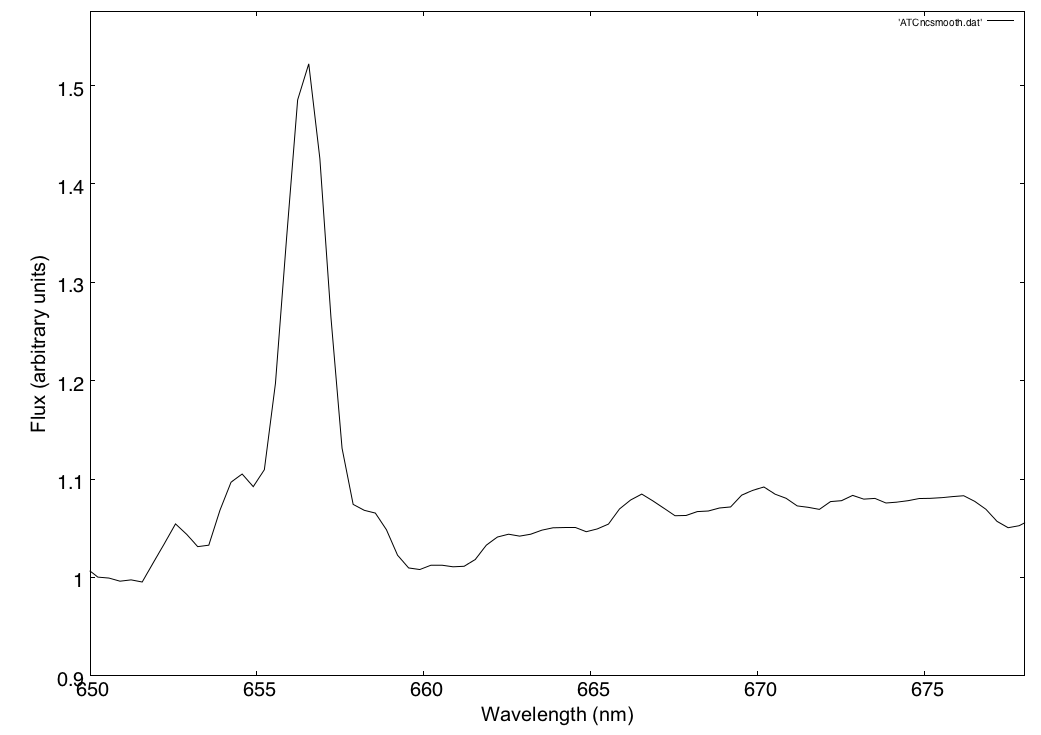}
    \caption{Spectrum of AT Cnc, smoothed with a 0.5-nm kernel.}
    \label{fig:ATspec}
\end{figure}

\subsection{A ``guest star" in Cancer in 1645 CE}

A ``guest star" star is mentioned in chapter 6 of the Korean encyclopedia
entitled "Chungbo Munhon Pigo" (= "revised encyclopedia"). This work,
which was compiled in 1907, was a revision of a work compiled in 1770.
The complete entry relating to the star gives only the year (23rd year
of King Injo) and lunar month -- the second month. The precise date is
not given; the Julian date is thus equivalent to some time between CE
1645 Feb 26 and Mar 27). The record simply states that "a large star
(daxing) entered (ru) Yugui." The duration of visibility is not specified.

The star group Yugui, in Cancer, consisted of the four stars theta,
eta, gamma and delta Cnc. The northernmost star of Yugui, gamma Cnc, is located at a Declination of +21 degrees 28', 
nearly 4 degrees south of AT Cnc. The brief text only mentions Yugui 
but does not use the expression xiu ("lunar lodge"). This suggests that the small and well-defined 
star group Yugui in Cancer, rather than the much larger lunar lodge, covering much of Cancer, and including AT Cnc, was intended.

The Chungbo Munhon Pigo is not a very reliable
source. For instance it falsely records "guest stars" in both AD 1600
and 1664 which, as comparison with the more detailed Sillok ("Veritable Records")  reveals,
were misplaced entries of the supernova of AD 1604 \citep{ 2002ISAA....5.....S} pp. 70-71. 
A similar error might have occurred around AD 1645.

To search for more potential information FRS has carefully checked
the Korean chronicle known as the Injo Sillok ("Veritable records of
the reign of King Injo"), a detailed day to day chronicle. It is mainly concerned
with non-astronomical matters but from time to time there are entries
of astronomical interest. A search of the chronicle for
the first and second months of the 23rd year of King Injo, reveals ten reports of astronomical events
-- such as conjunctions of the Moon with stars, eclipses, etc over
these two months. There is no reference to a guest star seen at Yugui.

An alternative interpretation is that the ``large star", said to ``enter"
Yugui, might have been a comet. However, there are no known records 
of either guest stars or comets in Chinese history for AD 1645. As this was only the second year of the
Qing dynasty after the defeat of the Ming in 1644, records may be incomplete in this turbulent period.

In summary, while the fragmentary Korean record of a transient in the southern part of the constellation Cancer in 1645 CE is
in intriguing proximity, both in time and location, to our deduced date of the last nova eruption of AT Cnc, the evidence currently available is 
insufficient to support a claim of a likely coincidence.




{99}

\bsp	
\label{lastpage}
\end{document}